\documentstyle[twocolumn,prl,aps,amssymb,psfig]{revtex}

\begin{document}
\twocolumn[\hsize\textwidth\columnwidth\hsize\csname@twocolumnfalse\endcsname
\draft

\title{Level-Spacing Distributions
       of Planar Quasiperiodic Tight-Binding Models}
\author{J.~X.~Zhong,$^{(1,2)}$ U.~Grimm,$^{(1)}$ R.~A.~R\"{o}mer,$^{(1)}$ 
        and M.~Schreiber$\,^{(1)}$} 
\address{
$^{(1)}$Institut f\"{u}r Physik, Technische Universit\"{a}t Chemnitz, 
D-09107 Chemnitz, Germany\\
$^{(2)}$Department of Physics, Xiangtan University,
Xiangtan 411105, P.~R.~China}

\date{$Revision: 1.13 $, printed \today}

\maketitle

\begin{abstract}%
  We study the statistical properties of energy spectra of
  two-dimensional quasiperiodic tight-binding models. We demonstrate
  that the nearest-neighbor level spacing distributions of these
  non-random systems are well described by random matrix theory.
  Properly taking into account the symmetries of models defined on
  various finite approximants of quasiperiodic tilings, we find that
  the underlying universal level-spacing distribution is given by the
  Wigner-Dyson distribution of the Gaussian orthogonal random matrix
  ensemble. Our data allow us to see the differences between the
  Wigner surmise and the exact level-spacing distribution.  In
  particular, our result differs from the critical level-spacing
  distribution computed at the metal-insulator transition in the
  three-dimensional Anderson model of disorder.
\end{abstract}

]
\narrowtext

Following the pioneering works of Wigner and Dyson \cite{WD}, random
matrix theory (RMT) has been successfully applied to investigate a
great variety of complex systems such as nuclear spectra, large atoms,
mesoscopic solids, and chaotic quantum billiards \cite{MH,E,BGS,C}.
In such systems, it has been shown that spectral fluctuations can be
modeled by universal level-spacing distributions (LSD) such as, e.g.,
$P_{\text{GOE}}(s)$ for the Gaussian orthogonal random matrix ensemble
(GOE) \cite{MH}.

A very natural application of RMT concerns disordered systems
\cite{AS}.  It has been shown that the metal-insulator transition
(MIT) in the three-dimensional (3D) Anderson model of localization is
accompanied by a transition of the LSD $P(s)$ \cite{SSSLS,HS,ZK}.
Here, $s$ denotes the energy spacing in units of the mean level
spacing $\Delta$.  In the metallic regime, $P(s)$ closely follows the
Wigner surmise $P_{\text{W}}(s)=\pi s\exp(-\pi s^2/4)/2$, which is a
good approximation of $P_{\text{GOE}}(s)$. On the insulating side,
$P(s)$ is given by Poisson's law $P_{\text{P}}(s)=\exp(-s)$.  One
important difference between the two distributions is their small-$s$
behavior: $P_{\text{W}}(s\rightarrow 0)\approx \pi s/2$ and $P_{\text
  P}(s\rightarrow 0)\approx 1$, indicating level repulsion and
clustering, respectively.  At the MIT, where the eigenstates are
multifractal \cite{SG}, another LSD, $P_{c}(s)$, has been observed
\cite{SSSLS,HS,ZK}.  

Multifractal eigenstates --- neither extended nor exponentially
localized --- have also been found in tight-binding (TB) models of
quasicrystals. In fact, these seem predominant in 1D and 2D \cite{FT};
in 3D, the attainable system sizes are yet too small for definite
statements \cite{RGS}. The multifractality is assumed to be connected to
the unusual transport properties of quasicrystals \cite{MB}, e.g.,
their resistivity increases considerably with decreasing temperature
and improving structural quality of the sample.  Thus, one may
speculate that the LSD in quasiperiodic models is also distinct from
the Wigner and Poisson behavior.

Quasicrystals lack the translational symmetry of periodic crystals,
but still retain long-range (orientational) order and show
non-crystallographic symmetries incompatible with lattice periodicity.
Thus, they constitute a class of materials somewhere in between
perfect crystals and amorphous systems.  Besides quasicrystals with
icosahedral symmetry \cite{SBGC}, which are aperiodic in any direction
of the 3D space, also dodecagonal \cite{INF}, decagonal \cite{B}, and
octagonal \cite{WCK} phases have been found, which can be viewed as
periodic stacks of quasiperiodic planes with 12-, 10-, and 8-fold
symmetry, respectively. Structure models of quasicrystals are based on
quasiperiodic tilings which can be constructed, e.g., by projection
from higher-dimensional periodic lattices \cite{J}.  We emphasize that
such quasiperiodic tilings, albeit yielding perfect rotationally
symmetric diffraction patterns, exhibit $n$-fold rotational symmetry
in a generalized sense only.  In particular, there need not be a point
with respect to which the tiling has an {\em exact}\/ global $n$-fold
rotational symmetry. If such a point exists, it is unique.

In order to understand the transport properties of quasicrystals
\cite{MB}, TB models defined on such aperiodic tilings (notably the
Penrose tiling) have received considerable attention
\cite{FT,confined,LM,octagonal,PSB,BS,PJ,ZY}.  For a TB model defined
on the octagonal (Ammann-Beenker) tiling \cite{AB}, the LSD has also
been used to classify the spectrum \cite{BS,PJ,ZY}.  For periodic
approximants, level repulsion was observed \cite{BS,PJ}, and $P(s)$
was argued to follow a log-normal distribution \cite{PJ}.  However, a
calculation for finite patches with an exact 8-fold symmetry yielded
level clustering \cite{ZY}.

In this Letter, we show that these somewhat diverging results become
comprehensible when one realizes that the tilings of
Refs.~\cite{BS,PJ} still retain non-trivial symmetries, such as a
reflection symmetry for the standard periodic approximants.  In order
to obtain the underlying universal LSD, one should consider the
irreducible subspectra separately, or break the symmetry by, e.g.,
either choosing patches without symmetry, or imposing suitable
boundary shapes as in quantum billiards, or introducing disorder.
Properly taking this into account, we find that the underlying LSD of
these non-random Hamiltonians is neither $P_{\text{P}}(s)$ \cite{ZY},
nor log-normal \cite{PJ}, nor $P_c(s)$ but rather $P_{\text{GOE}}(s)$.
The accuracy of our data further allows us to show that it is also not
$P_{\text W}(s)$ although this is as usual a reasonable approximation
\cite{MH}.  We emphasize that our results apply to all planar tilings
mentioned above.

\begin{figure}
\centerline{\psfig{figure=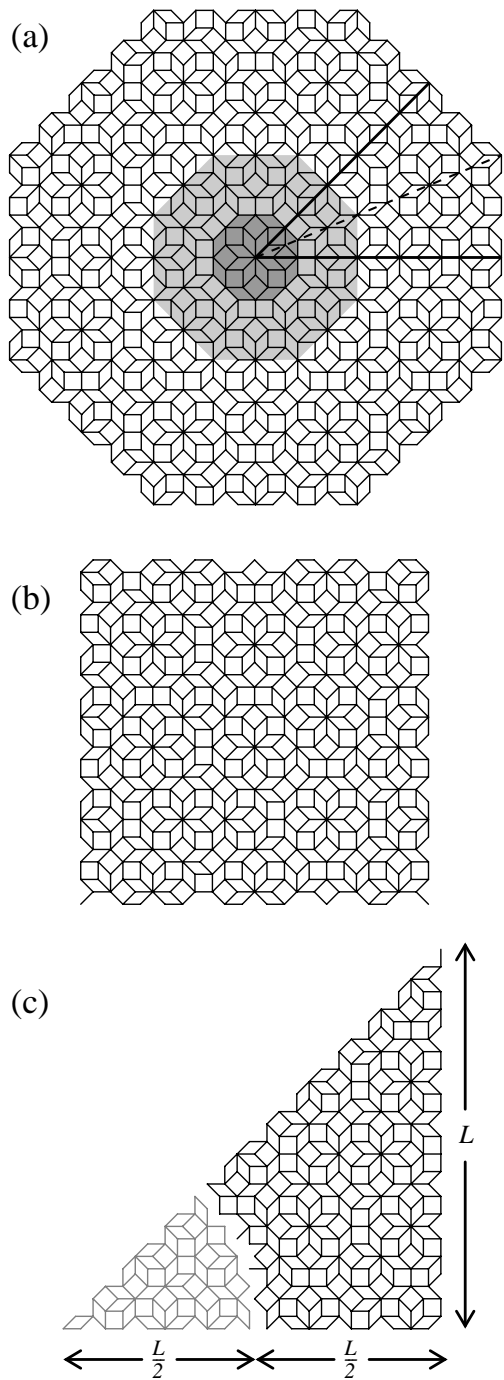,height=0.8\textheight}}
  \caption{
    (a)~Octagonal cluster of the Ammann-Beenker tiling with exact
    eightfold rotational symmetry around the central vertex
    $(x,y)=(0,0)$ as indicated by the solid and dashed lines. Shadings
    indicate successive inflation steps of the central octagon. The
    patch contains $833$ vertices.  (b)~Square-shaped cut defined by
    $0\leq x\leq L$, $-\case{L}{4}\leq y\leq \case{3L}{4}$ with $L=20$
    with 496 vertices.  (c)~Sinai billiard-shaped patch defined by
    $0\leq y\leq x\leq L$ and $x^2+y^2\geq \case{L^2}{4}$ with $L=22$
    with 246 vertices. The gray edges correspond to the interior of
    the circular arc; edges crossing the arc have been deleted.}
\label{fig1}
\end{figure}

Let us reconsider \cite{BS,PJ,ZY} the octagonal tiling consisting of
squares and rhombi with equal edge lengths as in Fig.~\ref{fig1}(a).
Besides the projection method mentioned above, one may also use the
self-similarity of the tiling to construct ever larger patches by
successive inflation steps \cite{K}.  E.g., the patch in
Fig.~\ref{fig1}(a) corresponds to two inflation steps of the inner
shaded octagon. On this tiling, we define the Hamiltonian
$H=\sum_{\langle i,j\rangle } |i\rangle \langle j|$ with free boundary
conditions, $|i\rangle$ denotes the Wannier state at vertex $i$, and
$\langle i,j\rangle$ runs over all pairs of vertices connected by an
edge of unit length.

We diagonalize the Hamiltonian by standard methods, and study the LSD
of the full spectrum. Due to the bipartiteness of the tiling, the
energy spectrum is symmetric about $E=0$. Furthermore, it has been
shown that a finite fraction of the states are degenerate at $E=0$
\cite{confined,PSB,BS,PJ}. These correspond to confined states
\cite{confined} limited to certain local environments, do not
contribute to the LSD, and we neglect them. In agreement with previous
calculations \cite{BS}, we find that the integrated density of states
(IDOS) is very smooth. This is different from 1D quasiperiodic systems
which typically have singular continuous spectra \cite{FT}.
Nevertheless, the IDOS is not strictly linear as required by RMT, so
we ``unfold'' the spectrum by fitting the IDOS to a cubic spline
\cite{HS} and use $s_{i}=N_{\text{av}}(E_{i+1})-N_{\text{av}}(E_{i})$
for the level-spacing at the $i$th level with $N_{\text{av}}$ the
smoothed IDOS. We remark that the unfolded LSD is not a bulk quantity
since $\Delta^{-1}$ is proportional to the system size. In what
follows, we shall always consider instead of $P(s)$ the integrated LSD
$I(s)=\int_s^\infty P(t)dt$ which is numerically more stable
\cite{HS,ZK}.

\begin{figure}
  \centerline{\psfig{figure=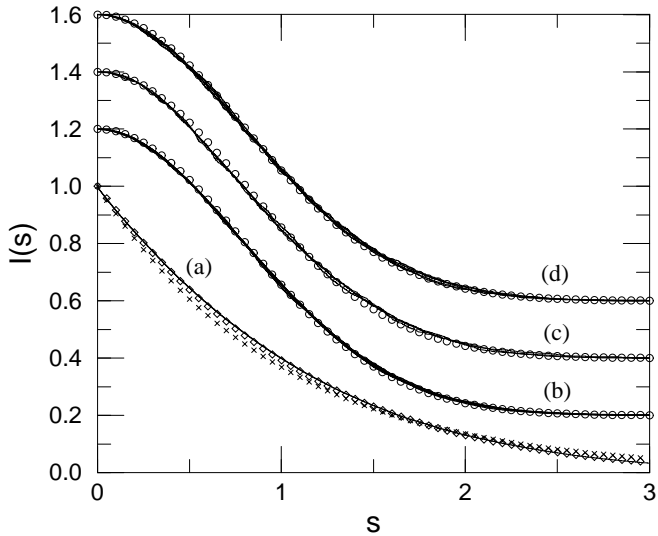,width=\columnwidth}}
  \caption{
  Integrated LSD $I(s)$ for
  (a)~the largest $D_8$-symmetric octagonal patch,
      crosses ($\times$) indicate $I_{\text P}(s)$, diamonds ($\diamond$)
      indicate $I_{\text W}^{(7)}(s)$; 
  (b)~the seven independent subspectra of the largest $D_8$-symmetric 
      octagonal patch, circles ($\circ$) indicate $I_{\text W}(s)$;
  (c)~squared-shaped patches of different sizes, circles as in (b);
  (d)~Sinai billiard-shaped patches of different sizes, circles as in (b).
  Curves (b)--(d) have been shifted by multiples of $0.2$ for clarity.}
\label{fig2}
\end{figure}

Fig.~\ref{fig2}(a) shows $I(s)$ obtained for a large octagonal patch
with 157369 vertices corresponding to three more inflation steps of
Fig.~\ref{fig1}(a).  At first glance, $I(s)$ seems to be close to the
integrated Poisson law $I_{\text{P}}(s)\equiv P_{\text{P}}(s)$, as
observed in Ref.~\cite{ZY}. However, this patch has the full
$D_8$-symmetry of the regular octagon, hence the Hamiltonian matrix
splits into ten blocks according to the irreducible representations of
the dihedral group $D_8$: using the rotational symmetry, one obtains
eight blocks, two of which split further under reflection, while the
remaining six form three pairs with identical spectra. This gives a
total of seven independent subspectra. As with the confined states, we
neglected the exact degeneracies induced by symmetry in
Fig.~\ref{fig2}(a), since they only contribute to $P(0)$. The
integrated LSD of the seven independent subspectra are shown in
Fig.~\ref{fig2}(b). Each subspectrum matches the integrated Wigner
surmise $I_{\text{W}}(s)=\exp(-\pi s^2/4)$ to very good accuracy.
{}From RMT it is known \cite{MH} that the LSD of a superposition of $k$
independent spectra, each of which obeys Wigner statistics, is given
by $P_{\text{W}}^{(k)}(s)={d^2\over ds^2}[\text{erfc}({\sqrt{\pi}\over
  2}{s\over k})]^k$ with $\text{erfc}(t)$ the complementary error
function.  For large $k$, $P_{\text{W}}^{(k)}(s)$ approaches the
Poisson law $P_{\text{P}}(s)$.  In Fig.~\ref{fig1}(a), we therefore
also included the integrated LSD $I_{\text{W}}^{(7)}(s)$ of $k=7$
Wigner spectra. The data clearly fits this curve better than $I_{\text
  P}(s)$. This explains why in a previous calculation \cite{ZY} a
Poisson-like distribution was found. We also note that our data do not
follow the integrated LSD $I_{c}(s)$ found at the Anderson MIT
\cite{ZK}.

\begin{figure}
  \centerline{\psfig{figure=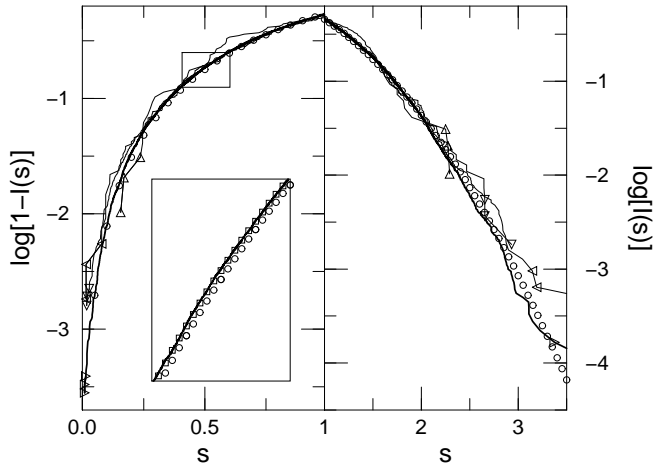,width=\columnwidth}}
  \caption{Small-$s$ (left) and large-$s$ (right) behavior of $I(s)$
           for one irreducible sector of $D_8$-symmetric octagonal
           patches of different sizes. The bold line corresponds to the
           largest patch. The three smallest and largest level spacings
           for each patch are denoted by triangles of different orientations.
           The circles ($\circ$) indicate $I_{\text W}(s)$.
           Inset: blow-up of the data
           region enclosed by the rectangle, showing only data
           for the largest patch. Squares ($\Box$)
           indicate $I_{\text{GOE}}(s)$. }
\label{fig3}
\end{figure}

Upon closer inspection of Fig.~\ref{fig2}(b), we see that there are
only very small differences between the seven integrated LSD, whereas
there are slightly larger deviations to $I_{\text{W}}(s)$. In
Fig.~\ref{fig3}, we show the small- and large-$s$ behavior in more
detail, restricting ourselves to data from one irreducible sector. We
include data for patches of different sizes, corresponding to two,
three, four, and five inflation steps of the inner shaded octagon of
Fig.~\ref{fig1}(a) with $833$, $4713$, $27137$, and $157369$ vertices,
respectively. The convergence to $I_{\text{W}}(s)$ with increasing
patch size is apparent both for small and large $s$ in
Fig.~\ref{fig3}.  However, the above-mentioned small deviations from
$I_{\text{W}}(s)$ still persist, even though the finite-size dependence
is already very small.  Therefore, we also consider in Fig.~\ref{fig3}
the {\em exact}\/ integrated LSD $I_{\text{GOE}}(s)$ \cite{MH}.
Although the Wigner surmise is usually a sufficient approximation of
the exact LSD, we show in the inset of Fig.~\ref{fig3} that our data
follow the exact curve $I_{\text{GOE}}(s)$.  Thus, the small
deviations seen in Fig.~\ref{fig2}, are due to the differences between
$I_{\text{W}}(s)$ and $I_{\text{GOE}}(s)$.

We can also approximate the octagonal tiling by patches which do not
have exact symmetries. For instance, in Fig.~\ref{fig1}(b) we show
such a square-shaped patch cut out of the octagonal tiling. Although
the quasiperiodic eightfold order is restored in the infinite patch,
there is never any exact symmetry present in the finite approximants.
The LSD is of the Wigner-type as shown in Fig.~\ref{fig2}(c) for
patches with side lengths $L=40$, $60$, and $80$, corresponding to
$1980$, $4392$, and $7785$ vertices, respectively. Thus, contrary to
the case of a simple square lattice exhibiting level clustering, we
find level repulsion.  Again, we observe small deviations from
$I_{\text{W}}(s)$ which can be explained as previously when using
$I_{\text{GOE}}(s)$.  If one uses square-shaped approximants with
symmetries, for instance squares centered around the eightfold point
of the patch in Fig.~\ref{fig1}(a) or the standard periodic
approximants used in Refs.~\cite{BS,PJ}, the LSD is again a
superposition of the LSD of the irreducible subspectra. Thus,
approaching the infinite tiling by square-shaped patches only slightly
shifted with respect to each other may give quite different LSD.  We
have obtained the same results also for circular patches. In this
case, one can have either $D_8$, or reflection, or no symmetry,
depending on the choice of the center.  Thus, the LSD is well
approximated by $P_{\text{W}}^{(7)}(s)$, or $P_{\text{W}}^{(2)}(s)$, or
$P_{\text{W}}^{(1)}(s)\equiv P_{\text{W}}(s)$, respectively.

A different way of excluding any symmetries is given by choosing
patches with special boundary shapes. In fact, this is well known in
the context of quantum billiards where it has been used to construct
quantum chaotic motion \cite{BGS,C}. One of the most prominent
examples is the Sinai billiard \cite{BGS,C}, which consists of $1/8$
of a square and a circular arc centered in the midpoint of the square.
Due to these boundary conditions, the LSD follows the Wigner surmise
even for free electrons \cite{BGS} instead of a Poisson law which is
found for integrable motion in simple square or circular billiards
\cite{MH}. In Fig.~\ref{fig1}(c), we show a Sinai billiard-shaped cut
of the octagonal tiling. Moving Sinai's billiard table across the
octagonal tiling, we can now generate many different patches. However,
in contrast to the square-shaped boundary, we never find a case that
retains any of the $D_8$-symmetries.  We computed $I(s)$ for
quasiperiodic billiards with $L=70$, $80$, $90$, $100$, and $110$,
corresponding to patches with $2416$, $3146$, $3969$, $4892$, and
$5905$ vertices, respectively.  The results presented in
Fig.~\ref{fig2}(d) follow $I_{\text{W}}(s)$, and, again, are even
closer to $I_{\text{GOE}}(s)$.

We emphasize that, apart from statistical fluctuations at small and
large values of $s$ as shown in Fig.~\ref{fig3}, there is no
systematic size-dependence of $I(s)$. This is in contrast to the 2D
Anderson model at weak disorder \cite{MMMS}, where a qualitative
change towards Poisson-like behavior for larger system sizes is
observed, indicating a finite localization length of the eigenstates.
The present size-independence of the LSD is compatible with
multifractal and extended states.

In conclusion, we have shown that the energy level statistics of TB
Hamiltonians defined on the octagonal tiling with different boundary
shapes is very well described by RMT. We can even see that our data
fit the exact integrated LSD $I_{\text{GOE}}(s)$ better than the
integrated Wigner surmise $I_{\text{W}}(s)$. This supports the
applicability of RMT for such completely deterministic Hamiltonians.
Although there is no randomness in these quasiperiodic models, one may
view the absence of translational symmetry as a sort of ``topological
disorder''. We clarify previous statements \cite{BS,PJ,ZY} and show
that the universal LSD for irreducible blocks of a symmetric patch, or
for patches without any symmetry is always $I_{\text{GOE}}(s)$. For
patches with the full $D_8$-symmetry, we find an integrated LSD which
is a superposition of seven $I_{\text{GOE}}(s)$.  Besides the
octagonal tiling, we have also considered planar 10- and 12-fold
quasiperiodic tilings and found analogous results.  In all these
cases, we never find a critical $I_{c}(s)$, distinct from
$I_{\text{GOE}}(s)$ and $I_{\text{P}}(s)$, as observed at the Anderson
MIT \cite{ZK}. This is somewhat surprising since eigenstates in these
quasiperiodic tilings are multifractal similarly to states at the MIT,
and we could have expected that this is reflected in the LSD.
Instead, we find that the LSD is similar to the LSD on the metallic
side of the MIT.

We thank M.\ Baake and I.\ K.\ Zharekeshev for discussions.  JXZ is
grateful for the kind hospitality in Chemnitz. Support from DFG (UG),
SFB393 (RAR), SEdC and the National Natural Science Foundation of
China (JXZ) is gratefully acknowledged.  We dedicate this work to
Hans-Ude Nissen, one of the co-discoverers of quasicrystals, on the
occasion of his 65th birthday.

\newpage

\end{document}